\documentclass[aps,pra,epsfigure,twocolumn,showpacs]{revtex4}
\usepackage{dcolumn}
\usepackage{bm}
\usepackage{graphicx}
\usepackage{amsmath}
\usepackage{latexsym}
\usepackage{amsfonts}
\usepackage{amssymb}
\usepackage{array}
\usepackage{epsfig}
\usepackage{times}
\usepackage{txfonts}
\usepackage{pifont}
\usepackage{amscd}
\usepackage{amsfonts}
\usepackage{textcomp}
\usepackage{hyperref}

\newcommand{\be}{\begin{equaion}}
\newcommand{\ee}{\end{equation}}
\newcommand{\bea}{\begin{eqnarray}}
\newcommand{\eea}{\end{eqnarray}}

\begin{document}
\title{Cavity Mode Frequencies and Strong Optomechanical Coupling in Two-Membrane Cavity Optomechanics}
\author{Jie Li$^1$, Andr\'e Xuereb$^{2,3}$, Nicola Malossi$^{1}$, and David Vitali$^{1}$}
\affiliation{$^1$School of Science and Technology, Physics Division, University of Camerino, I-62032 Camerino (MC), Italy, and INFN, Sezione di Perugia, Italy  \\
$^2$Department of Physics, University of Malta, Msida MSD 2080, Malta \\
$^3$Centre for Theoretical Atomic, Molecular and Optical Physics, School of Mathematics and Physics, Queen's University Belfast, Belfast BT7 1NN, United Kingdom}

\begin{abstract}
We study the cavity mode frequencies of a Fabry--P\'erot cavity containing two vibrating dielectric membranes. We derive the equations for the mode resonances and provide approximate analytical solutions for them as a function of the membrane positions, which act as an excellent approximation when the relative and center-of-mass position of the two membranes are much smaller than the cavity length. With these analytical solutions, one finds that extremely large optomechanical coupling of the membrane relative motion can be achieved in the limit of highly reflective membranes when the two membranes are placed very close to a resonance of the inner cavity formed by them.
We also study the cavity finesse of the system and verify that, under the conditions of large coupling, it is not appreciably affected by the presence of the two membranes. The achievable large values of the ratio between the optomechanical coupling and the cavity decay rate, $g/\kappa$, make this two-membrane system the simplest promising platform for implementing cavity optomechanics in the strong coupling regime.
\end{abstract}

\date{\today}
\maketitle

\section{Introduction}

Opto- and electro-mechanical systems in which a nanomechanical resonator is coupled to an optical or microwave cavity mode have been recently operated in the quantum regime by exploiting the so called linearised regime where the effective optomechanical interaction is enhanced by strongly driving the selected cavity mode~\cite{TeuflGSC,PainterGSC,KippenbergGSC,PainterSquee,PurdySquee,schwab,sillanpaa}. In this regime the system dynamics is linear and one is typically restricted to the manipulation and detection of Gaussian states of optical and mechanical modes~\cite{Hammerer2012}. However, there is a strong interest in realizing optomechanical devices able to reach the strong single-photon optomechanical coupling regime~\cite{favero,srinivasan,meenehan}, where the nonlinear nature of the radiation pressure coupling would allow the demonstration of novel phenomena. In fact, if the single-photon optomechanical coupling is large enough, the nonlinear dispersive nature of the radiation-pressure interaction would allow the observation of photon blockade~\cite{Rabl2011}, the generation of mechanical non-Gaussian steady states~\cite{Nunnenkamp2011,Xu2013}, nontrivial photon statistics in the presence of coherent driving~\cite{Liao2012,Xu2013b,Kronwald2013}, quantum non-demolition measurement~\cite{Ludwig2012}, single-photon detection~\cite{hong}, and quantum gates~\cite{Stannigel2012,asjad} at the single photon/phonon level. A further possibility is to use single photon optomechanical interferometry in this strong coupling regime for generating and detecting quantum superpositions at the macroscopic scale, eventually exploiting post-selection~\cite{Pepper2012,Vanner2013,Akram2013,Hong2013,Sekatski2014,Galland2014}.

The standard path for reaching the strong single-photon optomechanical coupling regime is to consider co-localised optical and vibrational modes~\cite{favero,srinivasan,meenehan}, with a large spatial overlap confined in very small volumes, corresponding to mechanical modes with extremely small effective mass. An alternative solution, capable of providing systems with a large ratio between the single-photon optomechanical coupling rate $g$ and the cavity decay rate $\kappa$, is to exploit quantum interference in multi-element optomechanical setups~\cite{andre,andre2}. In this case $g/\kappa$ can be increased by orders of magnitude even in more massive systems. Here we study in detail such a constructive interference enhancement in the simplest case of two parallel membranes within an optical cavity. We derive and solve the equation for the optical cavity mode resonance frequencies. The behaviour of these frequencies as a function of the center-of-mass (CoM) and relative distance of the two membranes provides a complete description of the optomechanical properties of the system and will allow us to establish which are the parameters to tune in order to reach large $g/\kappa$ values.

In such a two-membrane optomechanical system, the dependence of the cavity mode frequencies on the positions of the membranes is central to the description of the system, since it determines the optomechanical couplings~\cite{bhattacharya1}. However, we know that the mode equation is transcendental and cannot be solved analytically. The cavity resonance in such a system has been first studied in Ref.~\cite{bhattacharya2}, in which approximate analytical solutions of the mode equation are obtained in a perturbative manner. However, the solutions there are provided for only a few particular membrane positions, i.e., the equilibrium positions of the membranes are not left as free parameters in the optical frequencies. In this article, we instead provide approximate analytical solutions that work in more {\it general} situations, i.e., the optical mode frequency is a function of the CoM $Q$ and the relative position $q$ of the two membranes. With these analytical approximations, one can straightforwardly derive the optomechanical coupling for the CoM and the relative motion of the two membranes. We find that the optomechanical coupling of the latter can be significantly increased in the case of high-reflectivity membranes, $R_\text{m} \to 1$, when the two membranes are positioned such that the inner cavity they form is resonant. Such a coupling saturates to the value corresponding to the inner cavity, $g \propto \omega_0/q$ ($\omega_0$ is the cavity frequency) for very small $q$, as already shown in Refs.~\cite{andre,andre2}. These latter references focused on the scaling of the optomechanical coupling with the membranes at certain predefined fixed positions, without analyzing the generic dependence of the optical mode frequency versus the membrane positions along the cavity axis. Moreover they did not analyze in detail the effect of the membrane positions onto the cavity finesse. On the contrary, here we derive also an analytical expression for the cavity finesse versus the relative position $q$ of the two membranes. In particular, we have verified that the cavity finesse, and therefore the cavity decay rate, is not appreciably altered by the two membranes under the strong coupling condition; as a consequence $g/\kappa$ may be significantly increased, so that the two-membrane system is a promising candidate for the realisation of strong-coupling optomechanics. The present paper sheds new light on an experimentally-feasible instance of the optomechanical arrays studied in Refs.~\cite{andre,andre2}, which research it complements by providing analytical approximations to the properties and behaviour of the cavity around resonance.

The remainder of this paper is organized as follows. In Sec.~II we derive the exact equation for the cavity mode resonances in the presence of two membranes, we provide the approximate analytical solutions, and compare them with the numerical results. In Sec.~III we discuss the optomechanical coupling and provide approximate analytical formulas for such a coupling. Furthermore, we study the cavity finesse in the presence of the two membranes, especially in the large coupling regime. Finally, we reserve Sec.~IV for some concluding remarks.

\section{Cavity resonances}

\begin{figure}[b]
\includegraphics[width=0.9\linewidth]{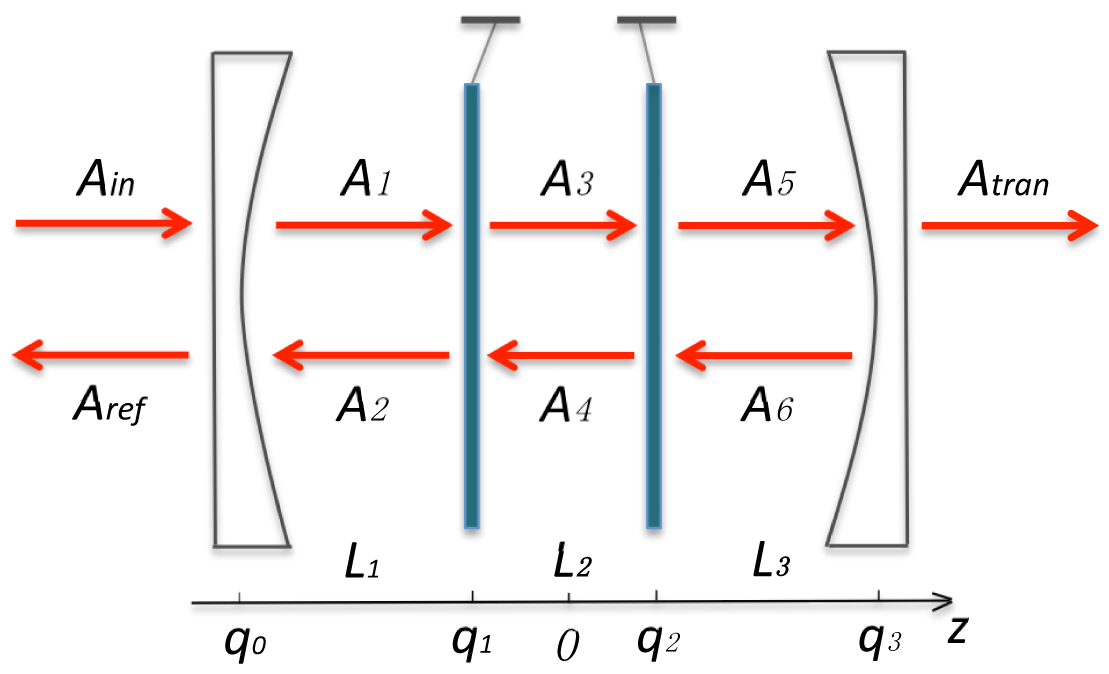}
\caption{Schematic diagram of the system: Two movable dielectric membranes are placed inside a Fabry--P\'erot cavity of length $L$ which is driven by an external laser. The position of two fixed mirrors (movable membranes) is denoted by $q_0$ and $q_3$ ($q_1$ and $q_2$); we have $L_i=q_i-q_{i-1}$ ($i=1,2,3$), with $q_{0,3}=\mp L/2$. }
\label{model}
\end{figure}

As shown in Fig.~\ref{model}, we consider two movable dielectric membranes placed inside a Fabry--P\'erot cavity with length $L$, which is driven by an external laser. The Fabry--P\'erot cavity is composed of two mirrors with electric field reflection and transmission coefficients $r_{1,2}$ and $t_{1,2}$. For simplicity, the cavity mirrors are assumed identical, i.e., $r\equiv r_{1,2}$ and $t\equiv t_{1,2}$; however, the results obtained in this paper can be extended in a straightforward way to the more general case of nonidentical mirrors. The reflection and transmission coefficients of a dielectric membrane of thickness $L_\text{m}$ and index of refraction $n$ are given by~\cite{Brookerbook}
\begin{equation}
\begin{split}
r_\text{m}&=\frac{(n^2-1)\sin\beta}{(n^2+1)\sin\beta+i\,2n\cos\beta},  \\
t_\text{m}&=\frac{2n}{(n^2+1)\sin\beta+i\,2n\cos\beta},
\end{split}
\label{rmtm}
\end{equation}
where $\beta=n k L_\text{m}$, and $k=2\pi/\lambda$ is the wavenumber of the electric field; $\lambda$ is its wavelength. In order to simplify our calculations, we assume that the membranes are identical.

The optical resonance frequencies correspond to the maxima of transmission of the whole cavity. The electric field amplitudes $A_j$ of incident ($j=\text{in}$), reflected ($j=\text{ref}$), and transmitted ($j=\text{tran}$) waves, as well as for the fields in the cavity ($j=1,2,\dots,6$), satisfy the following equations:
\begin{equation}
\begin{split}
A_1&=i\,t\,A_\text{in} + r\,A_2 e^{i k L_1},  \\
A_2&=i\,t_\text{m} A_4 e^{i k L_2} - r_\text{m} A_1 e^{i k L_1},  \\
A_3&=i\,t_\text{m} A_1 e^{i k L_1} - r_\text{m} A_4 e^{i k L_2},  \\
A_4&=i\,t_\text{m} A_6 e^{i k L_3} - r_\text{m} A_3 e^{i k L_2},  \\
A_5&=i\,t_\text{m} A_3 e^{i k L_2} - r_\text{m} A_6 e^{i k L_3},  \\
A_6&=r\,A_5 e^{i k L_3},  \\
A_\text{ref}&=i\,t\,A_2 e^{i k L_1} + r\,A_\text{in},  \\
A_\text{tran}&=i\,t\,A_5 e^{i k L_3},
\end{split}
\label{Aeqs}
\end{equation}
where $L_i$ ($i=1,2,3$) is the length of the subcavities formed by the mirrors and the membranes, i.e., $L_i=q_i-q_{i-1}$ ($i=1,2,3$), $q_{0,3}=\mp L/2$ (see Fig.~\ref{model}), so that $L=L_1+L_2+L_3$. We point the reader to Ref.~\cite{harris} for a similar approach in the case of a single membrane. The above equations, together with Eqs.~(\ref{rmtm}), are valid for any value of the thickness $L_\text{m}$, in the ideal one-dimensional case of plane waves, and flat and aligned mirrors and membranes. It can be applied also to the case of Gaussian cavity modes and spherical external mirrors as long as the membranes are placed within the Rayleigh range of the cavity. Membranes with very small absorption are available and therefore we will restrict to the case of real $n$, implying in particular ${\rm arg}(r_\text{m})={\rm arg}(t_\text{m})\equiv\phi$. 
Solving the above equations, the transmission ${\cal T}_\text{c}\equiv |t_\text{c}|^2=|A_\text{tran}/A_\text{in}|^2$ of the whole cavity is given by
\begin{equation}
{\cal T}_\text{c}=\frac{(1-R)^2(1-R_\text{m})^2}{|{\cal D}|^2},
\end{equation}
with
\begin{multline}
{\cal D}=1-R_\text{m} e^{i 2k L_2+i 2\phi}+R R_\text{m} e^{i 2k(L_1+L_3)+i 2\phi}-R e^{i 2k L+i 4\phi}\\+\sqrt{R R_\text{m}}\Bigl[e^{i 2k L_1+i \phi}+e^{i 2k L_3+i \phi}-e^{i 2k(L_1+L_2)+i 3\phi}\\-e^{i 2k(L_2+L_3)+i 3\phi}\Bigr].
\label{denom}
\end{multline}
We have taken $r=\sqrt{R}$, $t=\sqrt{1-R}$, and $R_\text{m}=\sqrt{r_\text{m}}e^{i \phi}$, $T_\text{m}=\sqrt{1-R_\text{m}}e^{i \phi}$, with $R$ and $R_\text{m}$ the reflectivity of the mirror and membrane, respectively. The external mirrors reflectivity will be taken as a given fixed parameter, which for typical high-finesse cavities is such that $1-R\sim10^{-5}$. For standard homogeneous membranes, the reflectivity $R_\text{m}$ associated with Eqs.~(\ref{rmtm}) takes values of the order of $0.1$--$0.4$, but patterned sub-wavelength grating membranes~\cite{Lawall} and photonic-crystal membranes~\cite{Bui,makles,groblacher,Deleglise} have been recently fabricated, and values up to $R_\text{m} \simeq 0.998$ have been achieved. Therefore $R_\text{m}$ will be taken as a variable parameter, eventually approaching $1$, but assuming in any case $R_\text{m} < R$. Re-expressing the quantities in terms of the relative motion $q=q_2-q_1$ and CoM coordinate $Q=(q_1+q_2)/2$,
after some algebra, the denominator in the transmission ${\cal T}_\text{c}$, i.e. $|{\cal D}|^2$, can be expressed in the following form
\begin{equation}
\begin{split}
|{\cal D}|^2&=A\,{\cal X}^2(k L') +B\,{\cal X}(k L') +C,
\end{split}
\label{denom2}
\end{equation}
where ${\cal X}(k L')\equiv\sin(kL')-R_\text{m}\sin(kL'-2kq')$, and $A, B, C$ are the coefficients given by
\begin{equation}
\begin{split}
A&=4R,  \\
B&=8\sqrt{R R_\text{m}}\,(1+R)\cos (2kQ) \sin (kq'),  \\
C&=8RR_\text{m} \cos(4kQ) \sin^2(kq')-2R(1-R_\text{m})^2  \\
&\,\,\,\,\,+(1+R^2)\left[1-2R_\text{m}\cos(2kq')+R_\text{m}^2\right].
\end{split}
\label{ABC}
\end{equation}
We have introduced the two parameters $L'\equiv L+2\phi/k$ and $q'\equiv q+\phi/k$, which can be considered as the effective cavity length and the effective membrane relative distance including the effect of the phase shift due to each membrane.

The equations derived in this section give access to the optical properties of a Fabry--P\'erot cavity with two identical membranes inside; we note in particular that the results of Refs.~\cite{andre,andre2} are limited to cavities with perfect end-mirrors (i.e., $R=1$). In what follows we will use the above expressions in experimentally-motivated limits to derive the optomechanical coupling strength for the relative motion of the two membranes.

\subsection{Derivation of the cavity mode resonance frequencies}

In the case of perfectly reflecting mirrors, $R=1$, the cavity mode resonances are given by the zeros of the denominator in the transmission ${\cal T}_\text{c}$, which in this case reduces to
\begin{equation}
|{\cal D}|^2= 4\left[{\cal X}(k L')+2\sqrt{R_\text{m}}\cos(2kQ)\sin(kq')\right]^2,
\end{equation}
so that the explicit equation for the cavity mode wavevector $k$ reads
\begin{multline}
\sin(kL')-R_\text{m}\sin(kL'-2kq')\\
+2\sqrt{R_\text{m}}\cos(2kQ)\sin(kq')=0.
\label{modeeq1}
\end{multline}
This expression is closely related to Eq.~(19) in Ref.~\cite{andre2}. In the general case $R<1$, the mode equation is obtained by minimizing the denominator $|{\cal D}|^2$. From Eq.~\eqref{denom2}, it is straightforward to see that when ${\cal X}(k L')=-B/2A$, i.e.
\begin{multline}
\sin(kL')-R_\text{m}\sin(kL'-2kq')\\
+\frac{1+R}{\sqrt{R}}\sqrt{R_\text{m}}\cos(2kQ)\sin(kq')=0,
\label{modeeq2}
\end{multline}
${\cal T}_\text{c}$ achieves its maximum value, that is
\begin{equation}
{\cal T}_\text{c}^\text{max}=\frac{(1-R_\text{m})^2}{(1-R_\text{m})^2+4R_\text{m} \sin^2(2kQ)\sin^2(kq')}.
\label{Tcmax}
\end{equation}
Eq.~\eqref{modeeq2} is therefore the exact equation for the cavity mode resonances, generalizing Eq.~\eqref{modeeq1} to the case $R\leq 1$. Eqs.~\eqref{modeeq1} and \eqref{modeeq2} cannot be solved analytically, but only numerically. However, in what follows, we show that excellent approximations of the analytical solution of Eqs.~\eqref{modeeq1} and \eqref{modeeq2} can be obtained under physically interesting conditions.
Eq.~\eqref{modeeq1} can be cast into the following form
\begin{equation}
{\cal A}(kq')\sin(kL')+{\cal B}(kq')\cos(kL')={\cal F}(kQ,kq'),
\label{ABF}
\end{equation}
where ${\cal A}(kq')=1-R_\text{m} \cos(2kq')$, ${\cal B}(kq')=R_\text{m} \sin(2kq')$, and ${\cal F}(kQ,kq')=-2\sqrt{R_\text{m}}\cos(2kQ)\sin(kq')$. We then divide both sides of Eq.~\eqref{ABF} by $\sqrt{{\cal A}^2+{\cal B}^2}$, and define ${\tilde O}=O/\sqrt{{\cal A}^2+{\cal B}^2}$, $O={\cal A}, {\cal B}, {\cal F}$. $|\tilde{\cal A}| \leq 1$ and $|\tilde{\cal B}| \leq 1$ by definition, while it is possible to explicitly verify that also $|\tilde{\cal F}| \leq 1$ holds. Therefore, we can rewrite Eq.~\eqref{ABF} in the equivalent form
\begin{equation}
\sin\left[kL'+\theta(kq')\right]=\tilde{\cal F}(kQ,kq'),
\label{kequation}
\end{equation}
where we have introduced the explicit dependence upon the variables $kq'$ and $kQ$,
\begin{equation}\label{efk}
    \tilde{\cal F}(kQ,kq')=-\frac{2\sqrt{R_\text{m}}\cos(2kQ)\sin(kq')}{\sqrt{1+R_\text{m}^2-2R_\text{m} \cos(2kq')}},
\end{equation}
and $\theta(kq')=(-1)^{{\rm Step}[-{\cal B}(kq')]} \arccos[\tilde{\cal A}(kq')]$, with ${\rm Step}(x)$ the unit-step function which is equal to $0$ for $x<0$ and to $1$ for $x\ge 0$. Note that since ${\cal A}(kq')>0$, one has that $\theta(kq')\in(-\frac{\pi}{2},\frac{\pi}{2})$.  The step function is introduced due to the fact that when ${\cal B}(kq')$ is positive, $\theta(kq')=\arccos[\tilde{\cal A}(kq')]\in (0,\frac{\pi}{2})$, while when ${\cal B}(kq')$ is negative, $\theta(kq')=-\arccos[\tilde{\cal A}(kq)]\in(-\frac{\pi}{2},0)$. Notice that Eq.~\eqref{kequation} is an equivalent form also for Eq.~\eqref{modeeq2} with an extremely good level of approximation, because $\frac{1+R}{\sqrt{R}}\simeq2$ for typical values of $R$.

\begin{figure*}[t]
\hskip0.15cm{(a)}\hskip6.63cm{(b)}\hskip5.53cm{(c)}
\includegraphics[width=0.33\linewidth]{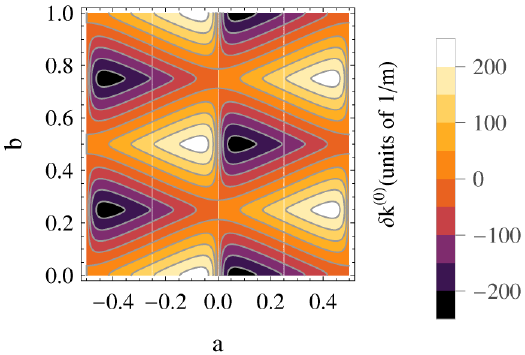}~~\includegraphics[width=0.33\linewidth]{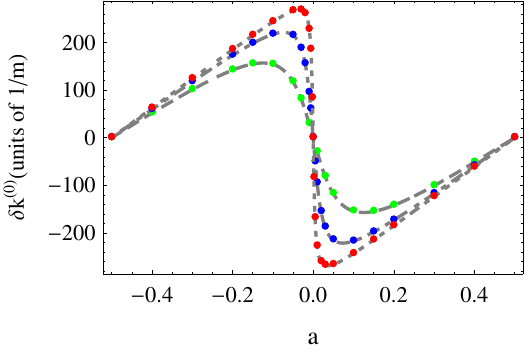}~~\includegraphics[width=0.318\linewidth]{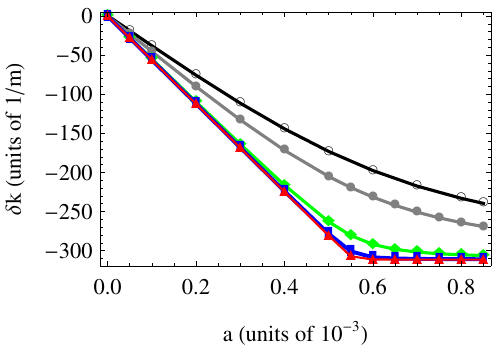}\\
\caption{(a)~Zeroth-order approximation $\delta k^{(0)}$ (in units of m$^{-1}$) as a function of $(q,Q)=(10.5\lambda+a\lambda,b\lambda)$ for $R_\text{m}=0.8$, with $a\in[-0.5,0.5]$, $b\in[0,1]$. (b)~$\delta k^{(0)}$ (curves) and exact numerical solution $\delta k$ (dots) versus $q=10.5\lambda+a\lambda$ ($Q=0$) for various values of the reflectivity: $R_\text{m}=0.5$ (dashed curve; green dots), $R_\text{m}=0.8$ (dot-dashed; blue dots) and  $R_\text{m}=0.95$ (dotted; red dots). (c)~Exact numerical $\delta k$ versus $q=10.5\lambda+a\lambda$ $(Q=0)$ with membrane reflectivity very close to the limit $R_\text{m}=1$. In practice we take $T_\text{m}=1-R_\text{m}=2\times10^{-3}$ (black); $10^{-3}$ (grey); $10^{-4}$ (green); $10^{-5}$ (blue); $10^{-6}$ (red). The rest of the parameters are $L=1$\,cm, $\lambda=1064$\,nm, $R=0.9999$, and $\phi=0$. Note that in (c) we consider unrealistic high reflectivity of the membranes $R_\text{m}>R$ in order to expose the saturation mechanism of the optomechanical coupling.}
\label{ultracoupling}
\end{figure*}

\begin{figure*}[t]
\hskip1.2cm{(a)}\hskip5.6cm{(b)}\hskip5.6cm{(c)}
\includegraphics[width=0.33\linewidth]{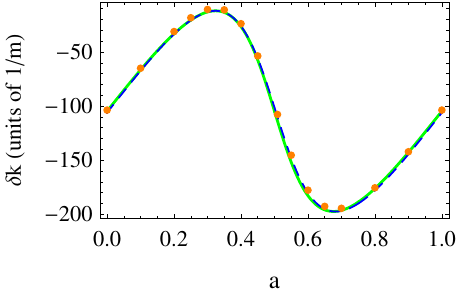}~~\includegraphics[width=0.33\linewidth]{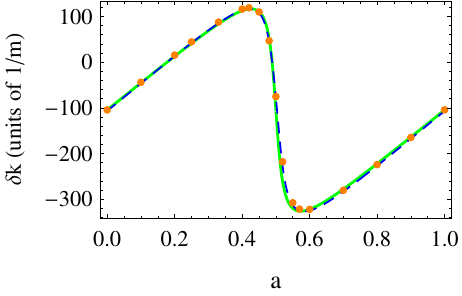}~~\includegraphics[width=0.33\linewidth]{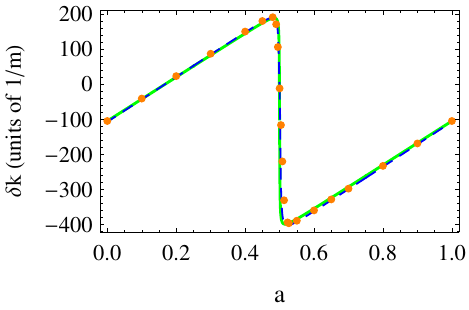}
\caption{Comparison of the zeroth- and first-order approximations $\delta k^{(0)}$ (green solid curve) and $\delta k^{(1)}$ (blue dashed) with the exact numerical solution of $\delta k$ (orange dots) versus $q=200\lambda-\frac{\phi}{k^{(0)}}+a\lambda$ ($Q=100\lambda$), for (a)~$R_\text{m}=0.2$, (b)~$R_\text{m}=0.8$, and (c)~$R_\text{m}=0.99$. We take $\phi=\frac{\pi}{6}$ and the other parameters are as in Fig.~\ref{ultracoupling}.}
\label{k0k1}
\end{figure*}

Eq.~\eqref{kequation} is equivalent to its formal solutions obtained by inverting the $\sin$ function,
\begin{equation}
kL'=m\pi+(-1)^m \arcsin[\tilde{\cal F}(kQ,kq')]-\theta(kq'),
\label{keq2}
\end{equation}
where $m = 1,2,3,\dots$. The case without membranes in the cavity corresponds to taking $R_\text{m}=0$, implying $\tilde{\cal F}(kQ,kq')=\theta(kq')=0$, when one obtains the standard empty cavity mode solutions $k_m^{(0)}=m\pi/L$. The insertion of the two membranes within the cavity is responsible for a frequency shift of each empty cavity mode, $k = k_m^{(0)}+ \delta k_m $. Since $k_m^{(0)}= 2 \pi/\lambda = m \pi/L$, and in typical experiments, $m$ is a very large integer because $\lambda \ll L$, this implies $ k_m^{(0)} \gg \delta k_m$, so that one can safely take $L'\simeq L+\frac{2\phi}{k_m^{(0)}}$ and $q'\simeq q+\frac{\phi}{k_m^{(0)}}$. Inserting the expressions of $k$, $L'$ and $q'$ into Eq.~\eqref{keq2}, the latter can be written as an equation for the frequency shifts alone,
\begin{multline}
\delta k_m=L^{-1}\Big\{ (-1)^m \arcsin[\tilde{\cal F}(k_m^{(0)}Q+\delta k_m Q,\, k_m^{(0)}q'+\delta k_m q')]\\
-\theta(k_m^{(0)}q'+\delta k_m q')-2\phi \Big\} \equiv h(k_m^{(0)}+ \delta k_m). \label{keq3}
\end{multline}
This equation is formally equivalent to the implicit equations for the cavity mode frequencies and wave vector Eqs.~(\ref{kequation}) and (\ref{keq2}), but it suggests a natural route for an approximate solution. In fact, we are looking for the frequency shift $\delta \omega_m = c \delta k_m$ around the optical frequency corresponding to the driving laser, $\omega_0  = c k_m^{(0)}$. Since $ k_m^{(0)} \gg \delta k_m$, it is reasonable to expand the right hand side of Eq.~(\ref{keq3}) as a Taylor series around $k_m^{(0)}$,
\begin{equation}
\delta k_m = h(k_m^{(0)})+ h'(k_m^{(0)})\delta k_m+\frac{1}{2}h''(k_m^{(0)})\delta k_m^2+\dots \, . \label{keq4}
\end{equation}
In what follows we drop the subscript $m$ whenever it is deemed unnecessary. It is possible to verify that the zeroth order solution $\delta k^{(0)} = h(k^{(0)})$ (see Fig.~\ref{ultracoupling}(a)) and the first order solution, $\delta k^{(1)} = h(k^{(0)})/[1-h'(k^{(0)})]$ provide a good approximate solution of the implicit equation Eq.~(\ref{keq3}) for not too large values of $q$ and $Q$, i.e., when $q/L, |Q/L| \ll 1$, and for values of $R_\text{m}$ not too close to 1. This is explicitly shown in Fig.~\ref{ultracoupling}(b) where the exact numerical solution of Eq.~(\ref{keq3}) is well reproduced by the zeroth order solution in the case $Q=0$ and $q/L \simeq 10^{-3}$, $ q \simeq 10 \lambda$. This is justified by the fact that one can rewrite
\begin{eqnarray}
h'(k)&=&\frac{Q}{L} d_Q+\frac{q}{L} d_q,\\
\label{d1hdk1}
h''(k)\,&=&\frac{1}{L}\,(Q^2\,d_{Q^2}+{q^2}\,d_{q^2}+2Qq\,d_{Qq})
\label{d2hdk2}
\end{eqnarray}
with $d_Q$, $d_q$, $d_{Q^2}$,$d_{q^2}$, $d_{Qq}$ dimensionless functions obtained by differentiating with respect to $kq$ and $kQ$. We have that $|d_Q|\leq 2$, while $d_q$, $d_{Q^2}$, $d_{q^2}$, and $d_{Qq}$ can be large, especially for highly reflective membranes, $R_\text{m} \to 1$, but nonetheless $h'(k)$ can be kept limited provided that $q/L, |Q/L| \ll 1$. This latter condition can be easily realised experimentally because one can always place the two membranes at the cavity center $Q=0$, and with a sufficiently small spacing between them, $q\ll L$, i.e., forming an inner cavity much shorter than the main one. Fig.~\ref{k0k1} shows that both the zeroth and first order approximations match quite well with the numerical solution of $\delta k$ even for larger values of $Q$ and $q$ when $R_\text{m}$ is not too close to unity, and the first order solution is slightly better than the zeroth order one when $R_\text{m}$ is large. From Figs.~\ref{ultracoupling} and \ref{k0k1}, we see that different choices of $\phi$ only shift the curves in $\delta k$ and $q$ axes without changing their pattern. In closing this section, we note that known results are mostly limited to the discussion of linear optomechanical coupling (however see Ref.~\cite{bhattacharya1} for a notable exception); the results presented in this section give access to coupling to higher powers of the displacement of the membranes and may in fact be straightforwardly extended to higher orders.

\section{Strong optomechanical coupling}

An important and evident aspect of Fig.~\ref{ultracoupling} is that it shows that it is possible to achieve \emph{strong single-photon optomechanical coupling} when the two-membrane system is placed at an appropriate configuration. In fact, Fig.~\ref{ultracoupling}(b) shows that a large single-photon optomechanical coupling with the relative motion, $g_q = c(\partial \delta k/\partial q) x_{\rm zpm}$ (with $x_{\rm zpm}=\sqrt{\hbar/M\omega_{\rm m}}$ the size of the zero-point motion of a mechanical resonator with mass $M$ and frequency $\omega_{\rm m}$) is achieved when $ q \ll L$ and $q+\frac{\phi}{k^{(0)}} \simeq p \pi/k^{(0)}$ (integer $p$), i.e., very close to a resonance of the inner cavity formed by the two membranes, especially in the limit $R_\text{m} \to 1$.

The possibility to enhance the optomechanical coupling with $N$ membranes within a Fabry--P\'erot cavity has been first pointed out in Refs.~\cite{andre,andre2}. Here we focus on the case of $N=2$ membranes in more detail, benefiting from our approximate analytical solutions of the cavity resonance presented in Sec.~II.
We derive the conditions under which one can achieve extremely large values of the derivative $\partial \delta k/\partial q\simeq \partial \delta k/\partial q'$ and therefore of $g_q$, by elaborating on Eq.~(\ref{keq3}) and on its zeroth order approximation, and we also derive simple analytical expressions for the dependence of $g_q$ upon $R_\text{m}$. We fix from now on the CoM coordinate $Q$ at a small value $Q \simeq 0$ and focus only upon the $q'$ dependence of $\delta k$. One can verify that $\delta k_m $ has the maxima and minima close to $ q'=2 p \pi/k_m^{(0)}$ (integer $p$) for $m$ even and at $ q'=(2 p+1) \pi/k_m^{(0)}$ for $m$ odd, and that the maximum shift is always bounded by $|\delta k|_{max}=2\pi/L$, which is approached for $R_\text{m} \to 1$. This is due to the fact that for the one-membrane case, the maximum frequency shift is $|\delta k|=\pi/L$ (corresponding to $R_\text{m}=1$), which occurs when the membrane is placed at the antinodes of the wave. Similarly, the same amount of frequency shift is induced by inserting the second membrane at the antinodes. Let us consider the case of odd $m$ in order to fix the ideas. Fig.~\ref{ultracoupling} shows that a large derivative $|\partial \delta k/\partial q'|$ ($q'=q$ when $\phi=0$) is achieved between two successive maxima and minima, at a value exactly given by $ q'=(2 p+1) \pi/k_m^{(0)}$. This fact, and the fact that $\tilde{\cal F}$ in Eq.~(\ref{efk}) is a function of $kq'$ only, suggest to write $k^{(0)}q'=(2 p+1) \pi+k^{(0)}\delta q' \equiv (2 p+1) \pi +\varepsilon$, and look at the behaviour of the shift $\delta k$ around $\varepsilon=0$. In fact, we expect that the maximum derivative and therefore the strongest optomechanical coupling, is achieved at a membrane distance $q$ smaller by $\frac{\phi}{k_m^{(0)}}$ from the inner cavity resonance condition $(2 p+1) \pi/k_m^{(0)}$.

After some algebra, we can rewrite also ${\cal A}(kq')$, ${\cal B}(kq')$, and $\tilde{\cal F}(kQ,kq')$ as a function of $\varepsilon$, obtaining
\begin{equation}
\begin{split}
{\cal A}(\varepsilon)&=T_\text{m}+2R_\text{m} \sin^2 \varepsilon, \\
{\cal B}(\varepsilon)&=R_\text{m} \sin (2\varepsilon), \\
\tilde{\cal F}(kQ,\varepsilon)&=\frac{2\sqrt{R_\text{m}} \cos (2kQ)\, \varepsilon}{\sqrt{T_\text{m}^2+4R_\text{m} \varepsilon^2}},
\label{bvare}
\end{split}
\end{equation}
where $T_\text{m}=1-R_\text{m}$. Using the zeroth order solution of the implicit equation Eq.~(\ref{keq3}), we then obtain the derivative of $\delta k$ with respect to $\varepsilon$. Neglecting high order terms of $\varepsilon$ in $\partial \delta k/\partial \varepsilon$, one then gets
\begin{equation}\label{shiftvare}
    \frac{\partial \delta k}{\partial \varepsilon}\simeq-\frac{1}{L} \,\frac{2\sqrt{R_\text{m}}}{T_\text{m}}\, \left[\cos (2k^{(0)} Q)+\sqrt{R_\text{m}}\right].
\end{equation}
As a consequence, one has that the single-photon coupling of the relative motion of the two membranes is given by
\begin{align}
g_{q}&=c \frac{\partial \delta k}{\partial q} x_{\rm zpm} \simeq c \left(\frac{\partial \delta k}{\partial \varepsilon} \frac{\partial \varepsilon}{\partial \delta q'} \right) x_{\rm zpm}   \\
&\simeq -\frac{\omega_0}{L} \frac{2\sqrt{R_\text{m}}}{T_\text{m}} \left[\cos (2k^{(0)} Q)+\sqrt{R_\text{m}}\right] x_{\rm zpm} \\
&=-\frac{\cos (2k^{(0)} Q)+\sqrt{R_\text{m}}}{T_\text{m}} g_{\rm sing},
\label{g_q}
\end{align}
corresponding to an enhancement by the factor $[\cos (2k^{(0)} Q)+\sqrt{R_\text{m}}]/T_\text{m}$ with respect to the maximum coupling of the single membrane case, $g_{\rm sing}=2\sqrt{R_\text{m}}(\omega_0/L)x_{\rm zpm}$. Therefore if $R_\text{m}$ is sufficiently close to 1, by placing the two membranes at the cavity center and with a carefully calibrated distance between them, one can achieve a strong single-photon coupling regime. Strong optomechanical coupling with the relative motion $q$ implies strong coupling with each membrane, because one has (for identical membranes) $g_1 = g_Q/2-g_q$ and $g_2 = g_Q/2+g_q$. Notice that there is no enhancement of the CoM coupling $g_Q$ (also refer to Fig.~\ref{ultracoupling}(a)).

However, Eq.~(\ref{g_q}) is valid when $R_\text{m}$ is not too close to unity and cannot be extended to the case of arbitrarily small $T_\text{m}$, i.e., one cannot achieve arbitrarily large coupling. In fact, this equation has been derived from the zeroth order solution for $\delta k$ which is no more valid when the first order term becomes relevant, i.e., when $|h'(k_0)| \simeq |(q/L)d_q| \lesssim 1$, which occurs just when $R_\text{m} \to 1$, when $d_q$ becomes very large. Using this fact, one has
\begin{equation}\label{condi}
  | g_q| =\, \left|\frac{\omega_0}{L} x_{\rm zpm}\, d_q\right| \,\leq \,\left|\frac{\omega_0}{L} x_{\rm zpm}\frac{L}{q}\right| = \frac{\omega_0}{q} x_{\rm zpm}=g_q^{\rm max},
\end{equation}
suggesting that the single-photon coupling can achieve at best the standard value corresponding to the small inner cavity of length $q$ formed by the two membranes, in the limit of highly reflective membranes $R_\text{m} \to 1$. This coincides with the results of Refs.~\cite{andre,andre2,wang} and it is also confirmed by Fig.~\ref{ultracoupling}(c), where the numerical solution of the implicit equation for the frequency shift for extremely small values of $T_\text{m}$ is shown. The saturation of the optomechanical coupling to a value which corresponds just to $g_q^{\rm max}$ of Eq.~(\ref{condi}) when $R_m \gtrsim 0.9999$ is evident.
Therefore, comparing with the expression for the single membrane case used in Eq.~(\ref{g_q}), one has that approaching the limit $R_m \to 1$, the single-photon optomechanical coupling rate is enhanced by an optimal double-membrane setup with respect to the single membrane case by the factor
\begin{equation}\label{enhanc}
  \left|\frac{ g_q^{\rm max}}{g_{\rm sing}}\right| =\frac{L}{2q}.
\end{equation}
Taking $L \sim 1$ cm for the cavity length and an achievable value $q \sim 10$ $\mu$m, which also guarantees that the high reflectivity of the membranes is not affected by near field effects, this corresponds to a significant increase by three orders of magnitude.

The physical argument at the basis of such a huge enhancement of the coupling when $R_\text{m} \to 1$ is that the optimal value for the membrane distance, $q \simeq p \pi/k^{(0)}-\frac{\phi}{k^{(0)}}$, corresponds to a field configuration in which the inner cavity formed by the two membranes is filled with a high intensity field, with a very weak field leaking out into the external cavity. In this case an infinitesimal change of the membrane distance corresponds to a big variation of the resonant frequency of the optical system and therefore to a large parametric radiation pressure coupling. In this regime one can achieve large coupling: the price to pay is that one needs an increasingly accurate control and stabilization of the membrane distance. In fact, it is possible to verify from the exact solution of Eq.~(\ref{keq3}) (see also Fig.~\ref{ultracoupling}(b)), that when $ R_\text{m} \to 1$, the interval of values for $q$ in which one has a very large coupling becomes narrower and narrower, and it scales to zero as $\lambda T_\text{m}/2\pi$. This scaling has not been discussed in previous treatments (cf., for example, Refs.~\cite{andre,andre2}) and emerges as a natural consequence of the analytical expressions obtained in this paper.

\subsection{Effects of the two-membrane system on the cavity finesse}

It is important to check the behaviour of the cavity finesse, and therefore of the cavity mode linewidth, in the configuration corresponding to the significant enhancement of the single-photon optomechanical coupling. In fact, strong optomechanical coupling means achieving a large ratio $g/\kappa$ which would also facilitate achieving large values of the single photon cooperativity $C_0 = g^2/(\kappa \gamma_\text{m})$, where $\gamma_\text{m}$ is the mechanical damping rate. Therefore we have to verify that $\kappa$ is not simultaneously increased when large coupling to the relative motion is established.

\begin{figure}[t]
\includegraphics[width=0.8\linewidth]{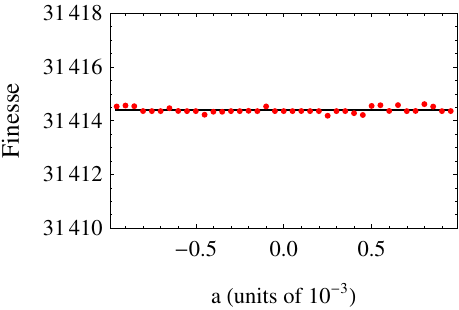}
\caption{The finesse ${\cal F}_{\rm cav}$ of the cavity in the presence of the two membranes (red dots) as a function of $q=10.5\lambda+a\lambda$ $(Q=0)$. The reflectivity of the membrane and of the mirrors are set equal to $R_\text{m}=0.999$ and $R=0.9999$, respectively, in correspondence with the grey curve in Fig.~\ref{ultracoupling}(c). The black line denotes the cavity finesse in the absence of the membranes, i.e., ${\cal F}_{\rm cav}=\pi\sqrt{R}/(1-R)$. Note that scatter of the data around the black line is due to numerical errors. The rest of the parameters are as in Fig.~\ref{ultracoupling}.}
\label{finessePlot}
\end{figure}

The cavity modes are obtained by solving the mode equation Eq.~\eqref{modeeq2}, with the optimal phase $\delta_m=kL' \equiv kL+2\phi$, which gives the maxima of the transmission ${\cal T}_\text{c}^\text{max}$. The transmission peaks can be approximated by a Lorentzian around the maxima, i.e., they can be written as a function of $\delta'=\delta-\delta_m$ for a given cavity mode, ${\cal T}_\text{c} \simeq \frac{\beta^2}{\beta^2+\delta'^2} {\cal T}_\text{c}^\text{max}$. The finesse of the cavity is related to $\beta$ by the relation ${\cal F}_{\rm cav}=\pi/(2\beta)$, and after tedious but straightforward calculations, one can see that it takes a relatively simple form when $Q=0$,
\begin{widetext}
\begin{equation}
{\cal F}_{\rm cav}=\frac{\pi \sqrt{ R R_\text{m}^2 \cos (2\delta_m-4k q') -2R R_\text{m} \cos (2\delta_m-2k q') +R \cos (2\delta_m) +(1+R)^2 R_\text{m} \sin^2 (k q') }}{(1-R)(1-R_\text{m})},
\end{equation}
\end{widetext}
which extends known results~\cite{andre2} to the domain of arbitrary membrane reflectivity and positions. In Fig.~\ref{finessePlot}, we compare the finesse of the cavity in the presence of the two membranes with that of the empty cavity without the membranes, ${\cal F}_{\rm cav}=\pi\sqrt{R}/(1-R)$, under the same conditions of Fig.~2 corresponding to an enhanced coupling $g_q$. We see that the finesse is not affected by the presence of the two membranes: this is an important result, showing that by placing the two membranes very close to each other and close to a resonance condition of the inner cavity formed by them, one can strongly enhance the single-photon optomechanical coupling $g_q$, while maintaining the same value of the cavity decay rate $\kappa$, since $\kappa=\pi c/(2L{\cal F}_{\rm cav})$. This result holds in the ideal situation we have assumed here of negligible absorption and scattering at the membranes. Recent experiments with high-reflectivity membranes~\cite{Lawall,Deleglise} have shown that optical absorption is actually negligible, but that scattering losses are responsible for a reduction of the cavity finesse. However, scattering losses can be mitigated and finesse reduction can become irrelevant provided that larger cavity mirrors are used. In any case, it is reasonable to assume that the cavity decay rate $\kappa$ will be essentially the same in the one and two-membrane case, so that using Eq.~(\ref{enhanc}), one has
\begin{equation}\label{enhanc2}
  \left|\frac{ \left(g/\kappa\right)_{\rm double}}{\left(g/\kappa\right)_{\rm sing}}\right| =\left|\frac{ g_q^{\rm max}}{g_{\rm sing}}\right|=\frac{L}{2q},
\end{equation}
that is, a significant increase, up to three orders of magnitude, of also the $g/\kappa$ ratio.

The explicit expression of the maximum value of such a ratio in the double-membrane case is given by
\begin{equation}
\frac{g_q^{\rm max}}{\kappa}=\frac{2\omega_0 {\cal F}_{\rm cav}}{\pi c} \frac{L}{q} x_{\rm zpm},
\label{goverk}
\end{equation}
which is achieved when the coupling $g_q$ saturates to its maximum value $g_q^{\rm max}$, which corresponds to $R_m \geq 0.9999$ with the parameters used in Fig.~\ref{ultracoupling}(c).
In this case, one reaches $g_q^{\rm max}/\kappa \simeq 1$ for the realistic set of parameters $L \simeq 1$ cm, $q \simeq 10$ $\mu$m, ${\cal F}_{\rm cav} \simeq 40000$, $M=2 $ ng, $\omega_{\rm m}=940 $ kHz. However, more importantly, for the recently achieved value of the membrane reflectivity
$R_m\simeq 0.998$~\cite{Lawall,Deleglise}, the numerical results of Fig.~\ref{ultracoupling}(c) show that $g_q \simeq 0.66 g_q^{\rm max}$, and therefore one can still achieve the strong single-photon coupling condition $g_q/\kappa \simeq 1$ by simply employing an external cavity with the higher value ${\cal F}_{\rm cav}\simeq 6\times 10^4$.
When combined with membrane vibrational modes with high mechanical quality factors (e.g., of the order of $10^6$), which has been recently shown to be compatible with high reflectivity membranes~\cite{groblacher}, this parameter regime corresponds to single photon cooperativities $C_0 \simeq 8\times 10^5$, significantly larger than the value $C_0 \simeq 8$ recently demonstrated by the single ``trampoline'' membrane-in-the-middle setup of Ref.~\cite{sankey}. In this parameter regime, many of the quantum nonlinear phenomena proposed in Refs.~\cite{Rabl2011,Nunnenkamp2011,Xu2013,Liao2012,Xu2013b,Kronwald2013,Ludwig2012,hong,Stannigel2012,asjad} could be demonstrated.

\section{Conclusions}

We have studied an optomechanical system of two vibrating dielectric membranes placed inside a Fabry--P\'erot cavity. We have derived the equation for the cavity mode resonance frequencies, and its zeroth and first order solutions that are excellent approximations of the implicit mode equation when the relative and CoM position of the two membranes, $q$ and $Q$, are much smaller than the cavity length. These analytical approximations provide a convenient tool to explore the rich physics of the system, and a full picture of the optomechanical coupling depending upon the position of the two membranes within the cavity. We stress that several of our expressions extend known results to the situation where the membranes are not tied to particular locations in the cavity (as opposed to Ref.~\cite{bhattacharya2}), and are more amenable to analysis and give access to further insight when compared to the generic $N$-membrane results first presented in Refs.~\cite{andre,andre2}.

We have shown, both numerically and analytically, that when the membrane reflectivity $R_\text{m}$ is close to $1$, very large single-photon optomechanical coupling of the relative motion is achievable when the inner cavity formed by the two membranes is close to resonance. We have also derived the analytical expression of the cavity finesse in the presence of the two membranes, and verified that, under the same conditions one has strong optomechanical coupling, the cavity finesse is not appreciably affected by the presence of the two membranes. As a consequence, one can achieve the single-photon strong coupling condition $g_q/\kappa \simeq 1$ when two high-reflectivity membranes with the recently demonstrated value $R_m = 0.998$~\cite{Lawall,Deleglise} form an inner cavity of length $q \simeq 10$ $\mu$m, placed in the middle of an external cavity of length $L \simeq 1$ cm and finesse ${\cal F}_{\rm cav}\simeq 6\times 10^4$.
This fact makes the two-membrane-in-the-middle system a very promising scheme for the implementation of the single-photon strong coupling regime of cavity optomechanics.

{\it Acknowledgments}.---We thank C.\ Genes for useful discussions and the anonymous referee for his/her detailed comments. A.\ X.\ thanks the University of Camerino for its kind hospitality. This work is supported by the European Commission through the Marie Curie ITN cQOM and FET-Open Project iQUOEMS.

\end{document}